\documentstyle[twoside,fleqn,espcrc2,epsf]{article}
\newcommand\be{\begin{equation}}
\newcommand\ee{\end{equation}}
\newcommand\bea{\begin{eqnarray}}
\newcommand\eea{\end{eqnarray}}
\def\psibar{{\bar \psi}}
\newcommand\half{{\textstyle{1\over2}}}

\newcommand{\AmS}{{\protect\the\textfont2
  A\kern-.1667em\lower.5ex\hbox{M}\kern-.125emS}}

\title{The Chiral Extension of Lattice QCD\thanks{This work is supported
in part by funds provided by D.O.E.  under contract \#DE-FG02-91ER40676.
Talk presented by R. C. Brower}}

\author{R. C. Brower\address{Department of Physics, 
        Boston University, 590 Commonwealth Ave., Boston, MA 02215},
        K. Orginos and C-I Tan\address{Department of Physics, 
        Brown University, Providence, RI 02912}}

\begin{document}

\begin{abstract}
The chiral extension of Quantum Chromodynamics (XQCD) adds to the
standard lattice action explicit pseudoscalar meson fields for the
chiral condensates.  With this action, it is feasible to do
simulations at the chiral limit with zero mass Goldstone modes. We
review the arguments for why this is expected to be in the same
universality class as the traditional action. We present preliminary
results on convergence of XQCD for naive fermions and on the
methodology for introducing counter terms to restore chiral symmetry
for Wilson fermions.
\end{abstract}
\maketitle

\section{INTRODUCTION}

In spite of the impressive progress in extracting useful physics from the
lattice formulation of Quantum Chromodynamics, in almost all cases  the
results rely on the quenched approximation. The additional cost of
simulating full QCD with internal fermion loops forces one to
unrealistically small lattices and heavy masses for the light
quarks. Even with the advent of Teraflops scale computing, new methods
will be needed to avoid these limitations.  Improved methods should be
sought (i) to allow accurate results on smaller lattices (i.e. larger
lattice spacings at a fixed volume) and (ii) to extrapolate reliably to
the chiral regime of light quarks.

Here we discuss a new lattice action (XQCD), which is intended to help
with the second problem (ii) by extending lattice QCD to include
explicit fields for the low mass Goldstone modes: the pion and eta. To
appreciate the concept behind this extension, it is useful to consider
it as a part of a more general approach that includes recent efforts to
address the first issue (i).

As originally suggested by Wilson in his renormalization group (RG)
approach, higher dimensional operators should be added to a lattice
action to suppress scaling violations due to the lattice spacing, $a$, or
the finite momentum cut-off, $\Lambda \equiv \pi/a$.  Recently this
program has gained new momentum because of the realization that even a
few higher dimensional operators can substantially reduce the lattice
artifacts.  For example, the clover action~\cite{clover} adds to the
five dimensional Wilson term, $\Lambda^{-1}\psibar D_\mu D_\mu \psi$,
the only other five dimensional operator,
\be
 \Lambda^{-1} \;\; \psibar \sigma_{\mu,\nu}  F_{\mu,\nu} \psi.
\label{eq:clover}
\ee
With the proper redefinition of the fermion fields, this allows one to
remove the $O(a)$ and $O(a g^{2n}_0 log^n(a))$ lattice artifacts in the
quark propagator. 

At dimension six, the  pure gauge operators, 
$$\Lambda^{-2} Tr[D_\mu F_{\mu,\nu} D_\lambda F_{\lambda,\nu}] \;\;\;\;
\Lambda^{-2} Tr[D_\lambda F_{\mu,\nu} D_\lambda F_{\mu,\nu}],$$
\be
\Lambda^{-2} Tr[D_\mu F_{\mu,\nu} D_\mu F_{\mu,\nu}],
\ee
when expressed in terms of the appropriate Wilson loops,
lead to improved gauge actions.  From the study of toy
models, there is also optimism that simple approximations to a
classically scale invariant (or ``perfect'') action can even more
dramatically reduce lattice artifacts, for asymptotically free
theories~\cite{perfect}.

At first sight our chiral extension of QCD appears to be a departure
from this trend in renormalization group improved actions.  However,
if we simultaneously block fermionic and gauge degrees of freedom, we
must bring in higher polynomials in the fermion fields as well.  For
example, in addition to several new gluon quark-bilinears similar to
Eq.~(\ref{eq:clover}), the remaining six dimension terms are 
local four-fermion operators. With two flavors, enforcing $U(2)
\times U(2)$ chiral-flavor invariance, there are 
only 4 possible four-fermion operators: the pion-eta operator,
$O_6(x)$,
\be
\Lambda^{-2}\;~[(\psibar \psi)^2 + (\psibar i\gamma_5 \psi)^2 
       +~(\psibar \vec{\tau} \psi)^2 + (\psibar  i\gamma_5
\vec{\tau}\psi)^2],
\label{eq:pieta}
\ee
and three others, $\Lambda^{-2} (\psibar \gamma_\mu \psi)^2$,
$\Lambda^{-2} (\psibar i \gamma_5 \gamma_\mu \psi)^2$, and $\Lambda^{-2}
[\;(\psibar \gamma_\mu \vec{\tau} \psi)^2 + (\psibar i \gamma_5
\gamma_\mu \vec{\tau} \psi)^2\;]$,
which are of less interest to us at present.

In its present form, XQCD amounts to a generalized lattice action
resulting from bosonizing the first of the four-fermion operators,
(\ref{eq:pieta}).  There are may other ways to motivate our
choice. For example in the standard discussions of the origins of
chiral symmetry breaking, one arrives at this same four-fermi
operator~(\ref{eq:pieta}) by the most attractive channel argument for
multiple gluon exchange graphs.  By introducing instanton effects, one
is led to the simpler four-fermion term of the Nambu-Jona-Lasino (NJL)
model,
\be
\Lambda^{-2} \; [(\psibar \psi)^2 + (\psibar  i\gamma_5 \vec{\tau} \psi)^2],
\label{eq:NJL}
\ee
which is a linear combination of $O_6(x)$ and the 'tHooft determinant.
The explicitly breaking chiral U(1) symmetry in NJL model removes the
need for an eta Goldstone mode.  Although we are studying XQCD for
both cases, Eqs~(\ref{eq:pieta}) and (\ref{eq:NJL}), it is probably
better to avoid introducing explicit chiral U(1) breaking at this
point, since on the lattice the anomaly is properly accounted for by
the doublers as  their masses go to infinity with the  cut-off.

Eventually, we may consider the complete set of six dimensional
operators using renormalization group methods to fix their
coefficients.  Indeed, several years ago~\cite{RCBcern}, we began with
the more ambitious goal of deducing XQCD by RG transformations of QCD
on a fine lattice. However, it is also useful to investigate the
immediate consequence of the simplest possible chiral extensions of
QCD in the same spirit as the limited investigations of separate improved
actions for the pure gluon and the Wilson fermion sectors.

\section{XQCD EFFECTIVE ACTION}

The standard lattice Lagrangian as formulated by Wilson is
\be
{\cal L}_{QCD}^{Lat} = \frac{2}{g^2_0 a^4} Re Tr[1 - U_{\mu,\nu}(x)] + \psibar
M(U) \psi
\label{eq:wilson}
\ee
with fermionic matrix,
\bea
&&\!\!\!\!\psibar M \psi = m_0 \psibar_x\psi_x  \\ 
&\!\!\!\!+&\!\!\!\!\frac{1}{2a}[\psibar_x \gamma_\mu U_\mu(x) \psi_{x+\mu}
- \psibar_{x+\mu} \gamma_\mu U^\dagger_\mu(x) \psi_x] \nonumber \\
&\!\!\!\!+&\!\!\!\!\frac{r}{2 a}[2 \psibar_x \psi_x -  \psi_x U_\mu(x) \psi_{x+\mu}
   - \psibar_{x+\mu}  U^\dagger_\mu(x)\psi_x ]. \nonumber
\eea
To simplify the present discussion of chiral invariance, we
drop the Wilson term ($r=0$) and ignore the doubling problem.

The Wilson QCD action has a UV fixed point at $g_0 = 0$. By tuning the
bare gauge coupling to the UV fixed point $g_0 \to 0$, the lattice
correlation length, $\xi$, will increase and eventually we enter the
scaling region where the lattice theory describes a continuum QCD
theory with renormalized coupling $g$. One may add our four-fermion
operator~(\ref{eq:pieta}) to the bare QCD action with an arbitrary
dimensionless coupling $G_0$,
\be
{\cal L}_{QCD}^{Lat} \to {\cal L}_{QCD}^{Lat} + G_0 O_6(x)~.
\label{eq:wilsonnjl}
\ee
According to the RG theory, up to $O(\xi^{-2}\ln^q\xi)$ terms, we can
always match the renormalized theory of the actions in
Eq.~(\ref{eq:wilson}) and Eq.~(\ref{eq:wilsonnjl}) such that the
effect of the four-fermion operator can be absorbed in a suitable
choice of the bare gauge coupling. In this sense the higher
dimensional operators are ``irrelevant'': the two actions belong to
the same universality class and their continuum limits are equivalent.

By a standard procedure, the four-fermion operator can be ``bosonized''
\be
{\cal L}_{QCD}^{Lat} \to {\cal L}_{QCD}^{Lat}
+ y_0 \psibar_x \widetilde \Phi_x \psi_x~+ \half \Lambda^2 Tr[\Phi^\dagger_x
\Phi_x],
\label{eq:ExQCD}
\ee
introducing a complex Lagrange multiplier field $\Phi_x = \phi_0 + i \vec
\tau \cdot \vec \phi$, where the tilde denotes the $\gamma_5$  projections,
\be
\widetilde\Phi = \frac{1+\gamma_5}{2}\Phi + \frac{1-\gamma_5}{2}\Phi^\dagger
\ee
and $y_0 = \sqrt{2G_0}$. It is very important to note that the ``mass''
parameter for the scalar field is $O(\Lambda)$, diverging with the cut-off.

Now we may imagine doing additional RG blocking transformations on the new
action. All operators that are absent in Eq.~(\ref{eq:ExQCD}) but
allowed under the chiral and gauge symmetries will be generated.  In
particular, both a kinetic term and a chirally invariant quartic term
will be generated for the $\Phi$ field,
\be
{\cal L}_{XQCD}^{Lat}={\cal L}_{QCD}^{Lat}+ {\cal L}_{\Phi} + y_0 
\; \psibar_x\widetilde\Phi_x \psi_x~,
\label{eq:latxq}
\ee
where 
\bea
{\cal L}_{\Phi} &=& - \half \kappa \sum_{\mu} Tr [\Phi^\dagger_x
 \Phi_{x+\mu} + \Phi^\dagger_{x+\mu} \Phi_x ] \nonumber \\
 &+&  V(Tr[\Phi^\dagger_x\Phi_x]).
\label{eq:latxqtwo}
\eea
It is this generalized scalar extension that we propose as a candidate action
for XQCD. The scalar field propagates with a mass at the cut-off so only short
range polynomial quark interactions are generated by its exchange.

On the lattice, it is also convenient to re-write the scalar field
$\Phi$ as $\sqrt{\Phi \Phi^\dagger} \Sigma$, where $\Sigma$ is an U(2)
matrix field. It is well known in lattice Higgs theories that one can
tune the potential so that the radial mode is frozen,
\be
\Phi \to \Sigma = e^{\textstyle i \eta/F_\eta} \;
e^{\textstyle i \vec \tau \cdot \vec \pi/F_\pi},
\ee
without changing the universality class.  We will do this to reduce the
number of new degrees of freedom.  We are studying two cases: U(2)
extended QCD with the eta field and SU(2) extended QCD without the eta
field.

\subsection{Induced Chiral Symmetry Breaking.}

XQCD is constructed from two chiral models: lattice QCD and the $\Sigma$
model coupled through the Yukawa term. Consequently for $y_0 \ne 0$
there is a single U(2) chiral symmetry and there must be
two phases -- a symmetric phase and a broken phase possessing the
Goldstone bosons with the quantum numbers of the $\pi$ and $\eta$. In
the broken phase the physical Goldstone modes are mixtures of the
``elementary'' $\Sigma$ and ``composite'' $(\psibar \gamma_5
\psi, \psibar i\gamma_5 \vec \tau \psi)$ chiral fields.

Here we give a qualitative survey of the phase diagram in the $(\beta,
\kappa, y_0)$ parameter space. Since $\Sigma^\dagger_x
\Sigma_x = 1$,  $y_0$ has units of mass.   For definiteness, we restrict the
discussion to the SU(2) version of XQCD ($det(\Sigma) =1$) without the
$\eta$ Goldstone mode.

For $y_0 = 0$, the scalar field sector, which is decoupled from the
QCD, is the $O(4)$ symmetric model with a critical point at
$\kappa_{cr}$. The broken phase ($\kappa > \kappa_{cr}$) has a
non-vanishing order parameter $\sigma \equiv \half Tr[\Sigma]$.  In
the symmetric phase (~$\kappa \le \kappa_{cr}$), the mass gap (or
sigma mass squared) is given by the difference,
\be
a^2 m^2_\sigma = 1/\kappa - 1/\kappa_{cr}.
\ee
We can anticipate that this is the appropriate region for regaining the
correct continuum limit of QCD, since we expect on the basis of our
earlier discussion that the mass squared parameter should be positive
(not tachyonic) and of the order of the cut-off ( $m_\sigma \sim \Lambda =
\pi/a$ ). 

\begin{figure}[ht] 
$$
\epsfxsize=7.0cm
\epsfysize=7.0cm
\epsfbox{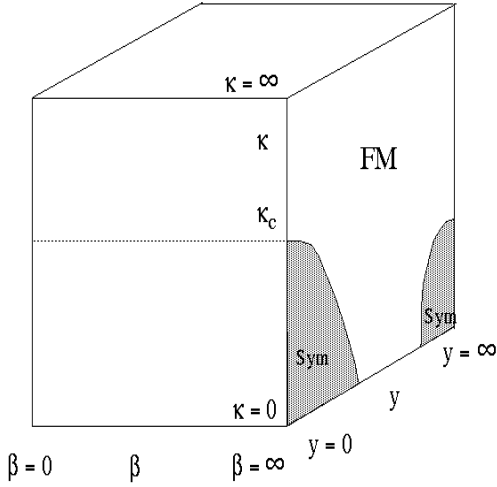}
$$
\caption{Phase diagram for XQCD. \label{fig:phasespace}}
\end{figure}

At $y_0 = 0$ we know that chiral symmetry in the QCD sector is broken
($<\psibar\psi>  \ne 0$) for all values of $\beta$. Therefore, as
indicated in Fig.~\ref{fig:phasespace} by the horizontal dotted line,
we have two phases in the $y_0=0$ plane:
\begin{eqnarray*}
<\sigma> &=&0, \;
\; <\psibar\psi> \ne 0, \; \; \kappa < \kappa_{cr}~, \\
<\sigma> &\ne& 0, \; \; <\psibar\psi> \ne 0, \; \;  \kappa >
\kappa_{cr}~.
\end{eqnarray*}
In the $g_0 = 0$ plane, Eq.~(8) becomes the $SU(2)
\times SU(2)$ symmetric Higgs-Yukawa model, which has been 
studied extensively~\cite{Bock}.  For our purpose, we are only concerned
with the critical surface in the weak $y_0$ region (see
Fig.~\ref{fig:phasespace}). The symmetric and broken phase is separated
by a second order phase transition line, $\kappa_{cr}(y_0)$, connected to $\kappa =
\kappa_{cr}$ at $y_0 = 0$.

When we couple the two models ($y_0 \ne 0$), it is important to
determine if the disorder of the $\Sigma$ sector can defeat the
chiral breaking mechanism of the standard QCD action. There are no
numerical results available in the region where $(\beta, \kappa, y_0)$
are all finite.  However an expansion in powers of $y_0$ can be easily
performed, since this is the only coupling
between the QCD and the $\Sigma$ sectors: $S_{XQCD} = S_{QCD} +
S_{\Sigma} + y_0 \sum_x \psibar_x \widetilde \Sigma_x \psi_x$.

To first nontrivial order in $y_0$, the order parameters $<\sigma>$ and
$<\psibar\psi>$ for chiral symmetry breaking are
\bea
\langle \sigma \rangle &=& \langle \sigma \rangle_\Sigma - y_0 \;
\chi_\sigma \; \langle \psibar\psi\rangle_{QCD} + O(y_0^2)~,
\label{eq:vev} \\
\langle\psibar\psi\rangle &=& \langle\psibar\psi\rangle_{QCD} - 
y_0 \langle\sigma\rangle_\Sigma \; \chi_{\psibar\psi} + O(y_0^2).
\label{eq:psibarpsi}
\eea
The VEV's  are ensemble averages in the QCD and $\Sigma$
sectors respectively,
\bea
\langle O \rangle_{QCD}&=& (1/Z_{QCD}) \; \int [dU d\psibar d\psi] 
\; O \; e^{-S_{QCD}} \nonumber\\
\langle O \rangle_\Sigma&=& (1/Z_{\Sigma})\; \int [d\Sigma ] 
\; O \; e^{-S_\Sigma}, \nonumber
\eea
and $\chi_{\psibar\psi}$ and $\chi_\sigma$ are the corresponding
susceptibilities.

Since we know that the QCD chiral symmetry is always broken
($\langle \psibar\psi \rangle_{QCD}
\ne 0~$) for $0 < \beta < \infty~$, it is clear from Eqs.~(\ref{eq:vev}) and
(\ref{eq:psibarpsi}) that XQCD must be in the broken phase $<\sigma> \ne
0, <\psibar\psi> \ne 0$ at any $\beta$ and $\kappa$ values in the small
$y_0$ region. Intuitively, this result is easy to understand. A
paramagnetic material will have nonzero magnetization in an external
magnetic field. For XQCD, the QCD sector acts as an external symmetry
breaking source to the scalar sector. Consequently, we are {\it always}
guaranteed to be in the correct chiral phase for continuum QCD. To
reach the continuum limit, we must approach the critical surface at
$g^2_0 =0$ without simultaneously tuning to the second order line at
$\kappa_{cr}(y_0)$.  This additional ``fine tuning'' would introduce an
unwanted finite mass scale for the Higgs field.

We have also explored the boundaries of the phase volume at the strong
gauge coupling plane $\beta = 0$ and at the strongly disordered chiral
plane $\kappa = 0$ to check that there is no sign of chiral symmetry
restoration.  We have carried out a perturbative
study for the weak coupling limit of XQCD near $\kappa_{cr}$. As a
result, we are confident that the only critical surfaces lie in the
weak coupling plane at $\beta^{-1} \sim g^2_0 = 0$ as depicted in
Fig.~\ref{fig:phasespace}.

\subsection{Toy Model Results}

XQCD resembles earlier chiral quark models, such as the Georgi-Manohar
(GM) model, in which  both quarks and explicit pions have been
introduced.  However, unlike these models, XQCD is not meant to be a
weak coupling effective theory. We do not assume new mass scales as
implied by the condition, $\Lambda_{\chi SB}<<\Lambda_{QCD}$, introduced
in the GM model. In XQCD, all ``spurious'' degrees of freedom
must have masses at the cut-off scale. To understand how these
properties might come about in a non-perturbative context, in a recent
paper~\cite{xqcd}, we have investigated several toy models in the
large N limit.

A natural toy model for XQCD is the Extended Nambu-Jona-Lasino model (XNJL),
which couples the sigma model to a NJL model via a chirally invariant Yukawa
term, $y_0\psibar(\sigma+i\gamma_5\vec\tau\cdot\vec\pi)\psi$.  Similar to our
expectation for XQCD, the physical sigma resonance and pion in this extended
model occur as a linear combination of elementary and composite operators:
$(\sigma, \vec \pi)$ and $(\psibar \psi,\psibar i\gamma_5\vec{\tau}\psi)$,
respectively.

We have demonstrated that the pseudoscalar mass spectrum is,
$$m_{\pi}^2=0 \;\; ; \;\; m_{\pi'}^2= \Lambda^2 \frac{y_0}{ g_0^2} 
(\frac{<\psibar \psi >}{<\sigma>}
+\frac{<\sigma>}{<\psibar \psi >}).$$
(Here, $g_0^2$ is the four-fermion coupling of the NJL model.)  There is only
one massless pion and the extra pion, $\pi'$, resides at the cut-off scale.
Similar results hold for the scalars. One mass, $m_{\sigma}$, which is
proportional to the VEV, remains in the physical spectrum, while the other
mass, $m_{\sigma'}\sim m_{\pi'}$, goes to infinity with the cut-off. Finally,
we have verified that as the Yukawa coupling, $y_0$, is tuned to zero the
spurious states remain at the cut-off. Just as we conjecture for XQCD, there
is a smooth decoupling limit, where only the physical pion remains in the
spectrum.  In the limit of infinite cut-off XNJL and NJL models (at least at
large N) are identical.

\subsection{Constituent Mass Extrapolation.}

\begin{figure}[ht]
$$
\epsfxsize=7.0cm
\epsfysize=7.0cm
\epsfbox{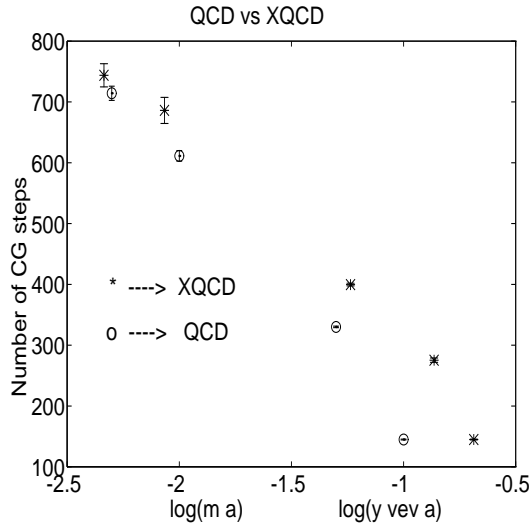}
$$
\caption{ Comparison of the convergence rate for the conjugate
gradient algorithm in QCD versus  XQCD.
\label{fig:convergence}}
\end{figure}

We have started to do numerical simulations of XQCD with naive
fermions using the $\phi$ algorithm~\cite{got}, weighting the
Trace-Log term appropriately for $n_f = 2$. The simulations were
performed on a 32-node CM5 using the CDPEAC routines of the BU-MIT
collaboration code.  For this preliminary investigation, the lattice
size was $8^4$ at $\beta=5.6$ and $\kappa=0.3$. For QCD, we varied the
fermion mass $m_0$ from $0.005$ to $0.1$, while for XQCD, we kept
$m_0=0.0$ and varied $y_0$ from $0.01$ to $1.0$. All runs started from
a cold Higgs field and a quenched gauge configuration at $\beta=5.8$.
After performing $100$ trajectories each with 30 MD steps, we measured
the number of iterations for the conjugate gradient (CG) algorithm to
converge to a residue of $10^{-11}$. The results are plotted in
Fig.~\ref{fig:convergence}.

One concludes that we can simulate naive fermions exactly at the chiral point,
$m_0=0$, without any convergence problem. Furthermore the convergence rate of
CG for XQCD depends on the effective constituent mass $y_0 <\sigma>$ of the
fermions and it exhibits the same convergence rate as QCD compared at $m_0
\simeq y_0 <\sigma>$. Larger lattices with a variety of values for
$(\beta,\kappa, y_0)$ are under investigation to better estimate the relative
efficiency between XQCD and QCD simulations.

The difficulty encountered with the standard QCD actions is the
necessity to extrapolate from large {\it bare} quarks masses for u and
d quarks ($\sim 50-200$ MeV) to the physical masses ( $\sim 5-10$
MeV).  In XQCD, we may consider an alternative extrapolation in the
{\it constituent} quark mass by taking the limit $y_0 \to
0$. Obviously in this limit, XQCD becomes the standard lattice theory.
In fact this approach is not so silly, if the limit is smooth and none
of the spurious state are re-introduced into the low mass spectrum as
$y_0\to 0$.

On the other hand, it should not be necessary to extrapolate to $y_0 =
0$. If we are in the same universality class, the fact that we regain
continuum QCD as the lattice spacing goes to zero means that $y_0$ is
an irrelevant parameter and the physical spectrum should be only weakly
dependent on the value of $y_0$ over a range of small values of $y_0$.
We are investigating numerically this range, but on intuitive grounds
any value of $y_0$ which contributes a relatively small part of the
300MeV constituent mass should be safe.

\section{DOUBLER PROBLEM} 

At first, it appears that a Wilson-Yukawa second order derivative term
would be the ideal choice for removing doublers, because it preserves
chiral invariance. At $g_0 = 0$, this leads to the Smit-Swift model
which was studied extensively in the last few years. There it was found
that a Wilson-Yukawa term either can not lift the doublers or removes
all fermions from the spectrum~\cite{Petcher}.

However, there is one difference which appeared at first to allow a new
solution. In our application, the VEV associated with the Higgs field can be
taken to the cut-off. We have studied this scenario in the context of the XNJL
toy model and found again that the spectrum looks fine, with one light
pseudoscalar meson and both the doublers and the spurious scalar states at the
cut-off. But there continues to be a fatal disease that looks to be very
general. Namely in spite of the separation in mass scales, the value of $F_\pi$
for the Goldstone mode remains at the cut-off. Consequently we are not
sanguine about this approach.

We are left with the options to use either the original Wilson
term~(\ref{eq:wilson}) or staggered fermions. The obvious problem with
both choices is their explicit breaking of chiral U(2) symmetry. Both
options require counter terms to restore (in the continuum limit) full
chiral symmetry.

Before considering Wilson fermions in more detail, we remark that
staggered fermions introduce very similar issues.  One can couple the
$\Sigma$ fields symmetrically to pairs of staggered flavors so that by
the usual trick of taking the square root of the fermionic determinant,
one is left with an SU(2) flavor form of XQCD. One nice feature of
staggered fermions is that the U(1) center of the chiral SU(2) group is
preserved so that vector Ward-Takahashi identities (WTI's) alone can be
used to restore the entire $SU(2) \times SU(2)$ current algebra.
Nonetheless, we feel that the price one pays for the flavor symmetry
breaking is still high and we prefer to use Wilson fermions.

\section{COUNTER TERMS} 

Since the Wilson term breaks chiral invariance, the Ward-Takahashi
identities (WTI's) associated with chiral symmetry are no longer
satisfied. One of these chiral Ward identities is the vanishing of the
pion two point function at zero external momentum $G_{\pi\pi}^{-1}(0) =
0$.  For the standard Wilson formulation, it can be shown that to
restore chiral symmetry in the continuum limit this is the only
condition that must be met. Thus the non-perturbative mass counter term
is introduced by tuning the bare quark mass $m_0 =
C_0 \; a^{-1} +  m_q$, where $m_0 = C_0 \; a^{-1}$ for a zero mass pion. (This is
usually expressed in terms of the Wilson hopping parameter, $K =
K_{crit} + O(a)$.)

\subsection{Axial Vector Current}

To restore chiral symmetry to XQCD, we must look at additional WTI's.
The full set of WTI's is found by local chiral rotation for the
integration variables in the path integral. For example, SU(2) chiral
invariance leads to
\be
< O \; \Delta_\mu A^a_\mu(x) > = < O \; \psibar \gamma_5 \tau_a ( m_0 + \chi_A) \psi >
\label{eq:axialwti}
\ee
where
\bea
A^a_\mu(x) &=& \half \psibar_x \tau_a \gamma_\mu \gamma_5  U_\mu(x) \psi_{x+\mu} \nonumber \\
&+& \half \kappa Tr[\{\tau_a,\Sigma^\dagger_x\} \Sigma_{x + \mu}]  - h. c.
\eea
For simplicity we have taken $O$ to be chiral invariant.  The $\chi_A$
term on the right comes from the variation of the Wilson term.  To
impose current conservation, we must introduce counter terms to cancel
the singular and constant contributions of $\chi_A$ in the limit of
zero lattice spacing.

As one may easily verify in lowest order perturbation theory, the
set of local counter terms required to {\it restore} chiral symmetry is:
\bea
V_{CT} &=&  \sum^4_{n=1} C_n Tr[(\Sigma)^n] + C_5 Tr[(\Sigma_{x +
\mu} - \Sigma_x)^2 ]   \nonumber  \\
&+& h.c. + A \; \psibar \psi  + B\; \bar \psi \widetilde\Sigma^\dagger
\psi.
\eea
There are additional counter terms for the wave functions and the
axial current, but these play a somewhat different role since they do
not have to be introduced into the bare action {\it prior} to the
simulations. They are only needed at the analysis stage.

\subsection{Non-Perturbative Counter Terms}

To fix the symmetry restoring counter terms, it is sufficient to
consider two matrix elements ( the vacuum and the single quark matrix
elements) of the WTI~(\ref{eq:axialwti}) with the $\Sigma_x$ field
treated as an external source.  Since the singular contributions only
occur for orders up to $y^4_0$, it is convenient to expand for small
$y_0$.  By expanding in $y_0$, the path integral factorizes between the
QCD and the $\Sigma$ sectors (see Section 2.1 ).

For example for the vacuum matrix element, one just computes term by
term (up to $y^4_0$) the vacuum expectation values of
\[
Tr[ \gamma_5 \tau_a \chi_A \frac{1}{M + \half y_0(\Sigma +
\Sigma^\dagger) + \half y_0 \gamma_5 (\Sigma - \Sigma^\dagger)}]
\]
in QCD at a fixed renormalized quark mass, $m_q = \mu$, computing the
necessary traces by using the standard method of averaging over Gaussian
pseudo fermionic sources. Each coefficient, $C_n(a y_0,g_0,a \mu,a)$, is then
expressed as a low order polynomial in $y_0$, so that once it has been
determined, XQCD is defined for a range of values of $y_0$ without
re-computing the counter terms.

\begin{table}[th]
\caption{One loop fermionic integrals.\label{tab:integral}}
\begin{center}
\begin{tabular}{|l|llll|} 
\hline 
$I_{n,m}$ &  m = 1 & m = 2 & m = 3  & m = 4 \\ \hline
n= 1   &0.2347 &0.8486   & 3.5421   & 16.318 \\
n= 2    &0.0259 &0.0597  &  0.1925 &0.7366 \\ 
n= 3    &(log-div)  & 0.0086  &0.0165  &0.0470 \\
n= 4    &(lin-div)  & (log-div) &0.0031  &0.0050  \\ \hline
\end{tabular}
\end{center}
\end{table}

We are in the process of computing the counter terms non-perturbatively
and comparing them with zeroth and first order expansions in $g^2_0$. 
To zeroth order, the single fermion loop contributes to $C_n, n
= 1,..,5$:
\begin{eqnarray*}
C_1 &=&  \frac{2}{a^3} y_0 \; I_{1,1} + \frac{2}{a} y^3_0 \; I_{3,3}  + O(g^2_0), \\
C_2 &=& - \frac{1}{a^2} y^2_0 \; I_{2,2} + 2 y^4_0 \;(I_{3,2} - I_{4,4}) +  O(g^2_0) , \\ 
C_3 &=&  \frac{2}{3 a} y^3_0 \;I_{3,3} +  O(g^2_0), \\
C_4 &=& - \frac{1}{2} y^4_0 \;I_{4,4} +  O(g^2_0)\\
C_5 &=&   y^2_0 \; I_Z +   O(g^2_0)
\end{eqnarray*}
The single loop integrals are,
\be
I_{n,m} = \int_{-\pi}^{\pi} \frac{d^4p}{(2\pi)^4}
\frac{[r w(p) + a \mu]^m} {[g(p)]^n},
\ee
with $g(p) = \sum_\mu sin^2(p_\mu) + (r w(p) + a \mu)^2$ and $w(p) = 
\sum_\mu (1 - cos(p_\mu))$. The values at $ \mu = 0$ and $r = 1$ 
are given in Table~\ref{tab:integral}. Since the Wilson term vanishes
at low momentum, no IR divergences occur in the chiral breaking terms
to this order. The expression for the wave function renormalization
integral, $I_Z$, is a little more complex but it  has the finite value
$I_Z = 0.05473$. The first contributions to $A$ and $B$ are
$O(g^2_0)$.

For U(2) extended QCD, we also need to impose the WTI for the U(1) axial
current. Due to the anomaly, this case is a little more intricate so we
postpone it to a future publication.

\section{CONCLUSIONS}

The chiral extension of QCD, provides an improved action in which one
is able to simulate near to or even at the chiral limit. Thus one can
study the chiral vacuum at the physical mass of the pion. To avoid
doublers, in the Wilson scheme, counter terms must be introduced, but
they can be computed numerically in standard QCD at the decoupling
point $y_0 = 0$. By studying the dependence of XQCD on the Yukawa
coupling one can test its insensitivity to $y_0$, an important signal
of universality. Also extrapolations in the constituent mass of the
quark, $y_0$, afford an alternative to the current practice of
extrapolating in the bare quark mass, $m_0$. We believe that being
able to invert the XQCD quark propagator in the chiral limit is an
important advantage of the chiral extension. Simulations are underway
to determine if this is true.

\end{document}